\input amstex
\loadmsbm
  \centerline{\bf A NEW APPROACH TO QUANTUM MEASUREMENT AND ENTANGLEMENT}
\centerline{Joseph F.\ Johnson, Math Dept., Villanova Univ.}
\vskip .7in
We introduce a formulation of combined systems in 
orthodox non-relativistic quantum mechanics, mathematically equivalent to the usual 
one.

This short suggestion seems to be new and could possibly lead to a more physically 
based understanding of entanglement.  It answers some of J. S. Bell's objections to 
the orthodox scheme of quantum mechanics, yet at the same time it interacts well 
with explicit analyses of the measurement process as an amplification process 
resulting from an externally driven phase transition.

{\bf 1. The reformulation}

If $V$ is the Hilbert space of wave functions of system 1, say for definiteness an 
electron, and if $W$ is the Hilbert space of wave functions of system 2, say 
for definiteness a distinguishable particle such as a meson, and if $H_1$ and 
$H_2$ are the Hamiltonian operators for each separate system, then the usual 
axioms of quantum mechanics almost force on us that the Hilbert space of states 
for the combined system is $V\otimes W$.  If the systems, although combined, have 
no interaction term in the Hamiltonian because they do not interact, then the 
Hamiltonian for the combined system is $H_1 \otimes I_2 + I_1 \otimes H_2$, where 
$I_1$ is the identity operator on $V$ and $I_2$ is the identity operator on $W$.
(From now on we write simply $I$ for whatever relevant identity operator should 
be understood from the context.)  (The sum is, of course, {\it not }\ the direct sum
but the tensor product is the outer tensor product).

More concretely, neglecting spin and so on, $V$ is an $L^2$ space on physical 
three-dimensional space, $L^2(\Bbb R ^3)$, as is $W$, and the natural concrete 
realisation of the tensor product is $L^2( \Bbb R^3 \oplus \Bbb R^3)$.
This is the usual one especially since it is so analogous with classical analytical 
mechanics of multi-particle systems.  It is also easily interpretable as a probability 
density on the corresponding classical configuration space $\Bbb R^3 \oplus \Bbb R^3$.
The value $\psi ((x_1,y_1,z_1)\oplus (x_2,y_2,z_2))$\ gives  the quantum amplitude 
that the respective distinguishable particles will be found at those two points 
of physical space.

But it is no longer a wave function on physical space, and this has seemed an 
obstacle to intuitive interpretations of the wave function in physical terms.
It has also been an obstacle to developing a relativistic many-body mechanics.

For this reason, J. S. Bell has registered objections to regarding the wave function 
as fundamentally physical and ultimate, and asked for a new physics which would 
be more closely associated with physical space even though this would even highlight 
the lack of locality of nature's laws.

But it is unnecessary to modify orthodox quantum mechanics in order to meet this 
objection.  The objection can be met with a simple mathematical equivalence.
The Hilbert space in question is obviously isomorphic to 
$$L^2(\Bbb R^3;W) = L^2(\Bbb R^3; L^2(\Bbb R^3)).$$
This isomorphism is functorial but also given quite easily by a simple formula such 
as
$$ \psi\otimes\varphi \mapsto w(x)$$
for $x\in\Bbb R^3$, a vector variable, and $w(x) = \psi(x) \cdot \varphi$, extended, 
of course, by linearity.

The inverse map is given by 
$$w(x) \mapsto 
\sum _i c_i(x) H_i(y) e^{-\vert\vert y \vert\vert^2}
$$
where $y
$ is the vector variable for the other copy of $\Bbb R^3$, the one underlying the 
concrete realisation of $W$, the $H_i$ are the various normalised Hermite polynomials 
in three dimensions, and the $c_i(x)$ are the Fourier coefficients of $w(x)$ in the 
basis we use here.

In this picture, the inner product on 
$L^2(\Bbb R^3;W)$ is given by $$\langle w(x),w'(x)\rangle = 
\sum_i <c_i,c'_i>.$$
Hence the isomorphism
$L^2(\Bbb R^3;W)= \bigoplus _{i=0}^\infty L^2(\Bbb R^3)$
respects the inner product structures implied.

Of course there is an equally valid isomorphism with $L^2(\Bbb R^3;V)$\ as well.
The lack of symmetry in this formulation is disturbing for some applications, 
but is less disturbing for the study of macroscopic amplification when $V$ is 
indeed as above but when $W$ is the state space of a large assembly of particles, 
approaching macroscopic dimensions.  The physics of such  a situation is already 
quite assymetric.  I propose that this formulation be used to study the situation 
in Wigner's formulation of the problem of quantum measurement, the entanglement 
of the incoming particle with the apparatus, after the measurement process is 
over.  We all believe now, and all the hard evidence supports this at greater 
and greater scales, that entanglement persists, even at lengths of dozens of 
miles (quantum teleportation) and at mesoscopic sizes, even macroscopic sizes.

{\bf 2. Observables on one factor alone}

Returning to the situation $V$, electron, $W$, meson.  Suppose the combined system 
is in the entangled state $\psi_1 \otimes \psi_2 + \varphi_1 \otimes \varphi_2$.
Suppose one makes an observation of system 2 in a way that would be represented 
by an observable on the second factor alone, $I \otimes Q$.  Here, $Q$ is the usual 
observable of some interesting quantity, such as energy or angular momentum, 
defined by operators on $W$ the same as if system 2 were in isolation.  At any 
rate,  $I \otimes Q$ is certainly hermitian if $Q$ is, and so the usual axioms 
of Quantum Mechanics say that this represents some physical aspect of the state 
of the combined system.  Suppose further that $\psi_2$ and $\varphi_2$ are 
normalised eigenvectors of $Q$ belonging to the eigenvalues 1 and -1, respectively.

How does the operator $I \otimes Q$  look in our reformulation?
It is easy to see that the expectation of $I \otimes Q$ on $w(x)$ is given by
$$\sum_{i,j}<c_i,c_j>\cdot Q_{i,j}$$
where $Q_{i,j}$ is the matrix coefficient of $Q$ with respect to the basis of $W$ 
we are using.  This can be expressed as, ``$Q$ acts on the values of $w(x)$.''

It has often been argued, but this is an ad hoc argument that does not follow 
from the axioms alone, that performing the measurement associated with $Q\otimes 
I$ should project the entangled state onto a statistical mixture.  (Now in fact, 
the concept of statistical mixture does not occur at all in the axioms, and should 
be banished from any discussion of the problem of quantum measurement that meets 
Bell's desiderata for being `serious'.)
We will examine the argument and translate it into our equivalent framework.

The usual argument proceeds as follows.  Let $c_1 = \vert \vert \psi_2\vert \vert 
^2$\ and 
$c_2 = \vert \vert \varphi_2\vert \vert ^2$.
These are the probability amplitudes that the result of a measurement of 
$Q\otimes I$ \ will yield 1, respectively -1.
After the measurement is made, the axiom of the reduction of the wave packet 
indicates that the combined system will be in the separable state 
$\psi_1 \otimes \psi_2$\ with probability $\vert \vert c_1 \vert \vert ^2$ 
but in the (equally separable) state  $ \varphi_1 \otimes \varphi_2$,
with probability $\vert \vert c_2  \vert \vert ^2$.


On the other hand, the new picture has a more intuitive description or characterisation 
of the sub-variety of all decomposable (separable) states.  A wave function of 
the combined system represents a decomposable state if and only if its values 
span a one-dimensional subspace of $W$.  

It follows from this that any operator of the form $I \otimes Q$ that causes a 
reduction of the wave function to a one-dimensional eigenspace of $W$ will, 
in our picture, cause a collapse onto the manifold of decomposable states.
(This is obvious in the usual picture as well.)

{\bf 3. Macroscopic observables on the second factor alone.}

The assymetry of our new picture is helpful in the context of quantum measurement.
In this context, $W$ is now the state space of a macroscopic body, a measurement 
apparatus.  There are various approaches to the idea of a pointer position.  A 
pointer position is, or should not be, quite the same as a linear closed subspace 
of states \dots  but whatever it is, should be some sort of approximation to the 
idea of a large but naked-eye indistinguishable collection of properties of the 
physical system underlying $W$.  A macroscopic observation of system 2 should 
approximate, in some sense, a projection of system 2 onto a one-dimensional subspace.
Therefore it induces a reduction of the wave packet of the combined system onto 
a decomposable state, satisfying ${\text im}
\Psi = \Bbb C \cdot w$, for some $w$\ an element of $W$.

But it is more interesting to consider more general operations on $W$ which, in one 
way or another, coarse-grain or obliterate thermodyamically somehow, its exact 
Hilbert space structure.  This is because two wave functions are as far apart as 
they can be from the standpoint of operators such as $Q$ if they are orthogoanl, 
yet some pairs of wave functions that are orthogonal would be, intuitively, 
naked-eye distinguishable, and some other pairs, indistinguishable.  So it seems 
to this authour unlikely that pointer position of system 2 can be adequately 
modelled by a closed linear subspace of $W$.
There are different rival suggestions as to how this would be done, but in this 
context the differences hardly matter, so consider any kind of collection of properties
$P_i$ of system 2 having the property that they are `physical,' they depend only 
on $W$ and not on $V$, and agree with our notion of pointer position.  

In arxiv.org/quant-ph/0502124, it was suggested that pointer position had to be 
modelled by a classical phase function on a discrete set $\Omega$\ of points, each point 
representing a stable thermodynamic limit (analogous to a phase) of system 2.
Before the measurement takes place, system 2 is in an unstable state, and which 
phase transition takes place depends on the microstate of system 1.  But this 
stimulus drives it into a stable phase, and any macroscopic observation of the 
`pointer position' yields the same answer for any microstate of system 2 within 
that phase, and hence for the purpose of macroscopic observations, the limit 
of system 2 is a classical zero-dimensional phase space, a finite set of 
disconnected points.  Since it is a thermodynamic limit, the space of states is 
the space of probability measures on the set, and since any positive measure becomes 
a probability measure after normalisation we may, in analogy with the use of $L^2(\Bbb
 R^3)$ instead of the manifold of normalised wave functions, take the space of states 
to be $\Cal M (\Omega)$ the space of positive atomic measures on $\Omega$.
Then, in the same approximation, the combined system 
becomes $$L^2(\Bbb R ^3, \Cal M (\{x_1, x_2, \dots x_n\}))
= \bigoplus^n_1 L^2(\Bbb R ^3). 
$$ 

There is a partial consensus, attacked by J. S. Bell and others, that a quantum 
mechanical observable $Q$ is macroscopic if its expectation on a superposition of 
states does not depend on the cross-terms.  Of course this only makes sense if we 
were to introduce some sort of notion of `realistic' superposition of states for 
which the phases were so distributed that this wash-out was approximately true. 
There would, as Bell points out, always be superpositions of states such that the 
phases were so unfortunately distributed that the cross terms led to significant 
deviations in the expectations of any given observable.  So this requires some sort 
of justification of attaching the adjective `typical' to some states of $W$ and 
not others, which has never been convincingly given.

This has been achieved for the thermodynamic limit of a sequence of $W$'s of greater 
and greater degrees of freedom.  (But in this case the limit is classical, which 
gets around the problem since $Q$ has been replaced by an autocorrelation coefficient
in the sense of a classical phase function.)  But the point of the present reformulation 
is in case more than one way of making precise the notion of macroscopic observable 
were to be found.  Whatever ways may be found, such as coarse-graining, there would 
be some notion of `indistinguishable' vectors $w$ and distinguishable ones.  There would 
be a physical basis to the approximation in terms of $W$ and the interaction with $V$,
and after the measurement process is over, this interaction would be effectively 
turned off with only the entanglement remaining, just as in the non-interacting picture.

In any case, whatever way of replacing $W$ by a `reduced description' of it, $\Omega$, 
is adopted, there will be a notion of $w,w' \in W$ being distinguishable in terms of 
the expectations of one (or another) macroscopic observable on the second factor alone
(corresponding to a reading of the pointer position) $Q$, and a notion of indistinguishable.
The points of $\Omega$ will then correspond to classes of macroscopically indistinguishable states.
$Q$ becomes essentially diagonal on this space.  Can an entangled state be distinguished 
from a decomposable state by means of expectations of macroscopic observables such as 
$Q$? 

Consider $\psi\otimes \varphi + \psi'\otimes \varphi'$.  There are two cases, 
when the second factors are distinguishable or not.  Suppose they are not.  
Then as far as expectations of $Q$ are concerned, we do not change the result by 
evaluating $Q$ on $(\psi + \psi')\otimes \varphi$, which is decomposable.  Suppose 
they are.  Then, (since the `cross-terms' wash out), $<Q>$ becomes the same as 
the expectation of $Q$ on the classical mixed state (probability distribution) on 
$\Omega$, $\vert\vert c\vert\vert ^2 \chi_i+\vert\vert d\vert\vert^2\chi_j$ (where 
$\chi_i$ is the characteristic function on the reduced phase description of the class 
of $\varphi$, etc., and $c(x)$, respectively, $d(x)$, are the reformulations of 
$\psi$, respectively, $\psi'$.  But this in turn has the same expectation as the 
decomposable state

$$ \frac\psi{\vert\vert \psi\vert\vert } \otimes \{
\vert\vert \psi\vert\vert^2 \varphi +
\vert\vert \psi' \vert\vert^2 \varphi'
\},$$ independently of $Q$ or of whether the states are distinguishable or not.

So the phenomenon of entanglement simply cannot be detected by macroscopic pointer 
variables.  
\end